\documentclass[a4paper,twocolumn,twoside,fleqn,10pt]{article}
\usepackage{amsmath}
\usepackage{authblk}
\usepackage[svgnames]{xcolor}
\definecolor{blue}{RGB}{0,0,255}
\usepackage{caption}
\usepackage[font={color=blue},figurename=Fig.,labelfont={it}]{caption}
\usepackage{bm}
\usepackage{subcaption}
\usepackage[varg]{txfonts}
\usepackage{graphicx}
\usepackage{pdfpages}
\usepackage{wasysym}
\usepackage{rotating}
\usepackage{hyperref}
\usepackage{indentfirst}
\usepackage{float}
\usepackage{lipsum} 
\usepackage{setspace} 
\usepackage{indentfirst}
\usepackage[english]{babel}
\usepackage[switch,columnwise]{lineno}
\usepackage[left=2cm,right=2cm,top=2cm,bottom=2cm]{geometry}

\hypersetup{
    colorlinks=true,                          
    linkcolor=blue, 
    citecolor=blue, 
    urlcolor=blue,
    linktoc=all
}

\newcommand{\blue}[1]{\textcolor{blue}{#1}}

\newcommand{\eg}{\textit{eg }}
\newcommand{\cf}{\textit{cf }}
\newcommand{\ie}{\textit{ie }}

\newcommand{\Lagrange}{\blue{Lagrange} }
\newcommand{\Laplace}{\blue{Laplace} }
\newcommand{\LeMouel}{\blue{Le Mouël} }
\newcommand{\Newton}{\blue{Newton} }
\newcommand{\Kepler}{\blue{Kepler} }
\newcommand{\Einstein}{\blue{Einstein} }
\newcommand{\Poincare}{\blue{Poincaré} }

\usepackage{fancyhdr}
\begin{document}
\title{On distributions law of planetary rotations and revolutions as a function of aphelia, following Lagrange’s formulation.}
\author[1]{J-L. Le Mouël}
\author[1]{F. Lopes\thanks{lopesf@ipgp.fr}}
\author[1]{V. Courtillot}
\author[2]{D. Giber}
\author[3]{J-B. Boulé}
			
\affil[1]{Universit\'e Paris Cité, Institut de Physique du globe de Paris, CNRS UMR 7154, F-75005 Paris, France}
\affil[2]{LGL-TPE - Laboratoire de Géologie de Lyon - Terre, Planètes, Environnement, Lyon, France}
\affil[3]{CNRS UMR7196, INSERM U1154, Museum National d'Histoire Naturelle, Paris, F-75005,  France}

\date{\today}
\maketitle

\abstract {We explore the links between the periods of rotation and revolution of planets, following Lagrange’s presentation of mechanics Lagrange (1788). The energy of a planet in motion in a central field is the sum of kinetic, centrifugal and centripetal energies. For each planet, one can calculate a “constant of gravitation” $\mathcal{G}_p$. For the giant planets, $\mathcal{G}_p$ decreases as a function of aphelia $a$. There is no such organized behavior for the terrestrial planets. The perturbing potential of other planets $\dfrac{\gamma}{r^3}$ generates a small angular contribution to the displacement , identical to Einstein’s formula for precession. Delays in the planetary perihelia follow a (-5/2) power law of $a$. The differences in delays are negligible from Mars to Neptune. For the three telluric planets the situation is different. This is readily understood in the Lagrange formalism. The telluric planets have lost energy, transferred to the planetary rotations via inclination of their axes, as predicted by Lagrange’s top. The ratio of areal velocities to rotation obeys a (-5/2) power law of $a$. The ratio of areal velocity to integrated period R also fits a (-5/2) power dependence, implying linearity of the energy exchange between revolution and rotation. The perihelion delays, the areal velocities and the planetary rotations display power laws of aphelia, whose behavior contrasts with that of the kinetic moment. The areal velocity being linearly linked to the kinetic moment of planets, this must be the level at which the transfer is achieved. The law of rotation periods as a function of aphelia gives the variations of inclination of the rotation axis. Since one is in a closed system, this ratio should be constant for each planet: all planets do follow the same power law of aphelia. This must be the reason for the presence of commensurable periods of the four Jovian planets, taken alone, in pairs and in pairs of pairs, in many geophysical and climatic series.
}

\maketitle

\section{\label{sec1} Introduction}
	In this paper, we explore the possibility of uncovering general planetary laws that link the rotation, revolution and rotation axes of planets. We start with a 'memento' of some of the main results produced by \Lagrange (\cite{Lagrange1788}), following his formulation of mechanics. We derive some properties of distributions of key variables of planetary orbits as a function of aphelia or distance to the Sun. Finally, we discuss the importance and potential use of these results. \Laplace (\cite{Laplace1799}) has shown that a planet's rotation and the inclination of its rotation axis are connected. Under the action of planetary torques, the inclination of the rotation axis reacts as does a weight acting on a rotating top, resulting in a transfer of energy.

\section{\label{sec2} A memento of Lagrangian mechanics}
We first start with a summary of the basic equations of mechanics, a summary that can be found in most physics graduate textbooks, involving gravitational potential, planetary torques and centrifugal forces, as elegantly presented by \Lagrange  (\cite{Lagrange1788}). \\

	In a Galilean reference system, in which time is uniform and the physical laws are homogeneous and isotropic, any mass in motion in that reference system conserves three quantities (primary integrals): energy ($\mathcal{E}$), impetus ($\mathcal{P}$) and moment ($\mathcal{M}$). Let a planet with mass $m$ and cylindrical coordinates ($r,\varphi$) be in motion in a central field $\mathcal{U}(r)$ of the form $-\dfrac{\alpha}{r}$ (always an attractive one \cite{Newton1687,dAlembert1749,Chastelet1756}). The symmetry axis is labeled $z$. As implied by the law of transformation of the kinetic moment, mechanical properties of the system do not change under any rotation about this axis. The moment must be defined with respect to a point located on that same axis. The total energy of the system can be written as:
\begin{subequations}
    \begin{equation}
        {\bf{\mathcal{E}}} = \dfrac{m}{2}(\dot{r}^{2}+r^{2} \dot{\varphi}^{2}) + \mathcal{U}(r) = \dfrac{m\dot{r}^{2}}{2}+\dfrac{\mathcal{M}^{2}}{2mr^{2}} +\mathcal{U}(r)  
        \label{eq:01a}
    \end{equation}
	\label{eq:01}
	that is the sum of the kinetic, centrifugal and centripetal energies. This formulation is more complete than \Newton ’s own (\cite{Newton1687}) and was the basis for the main criticism addressed to the scientist over the years (\eg \cite{Laplace1799,dAlembert1749,Chastelet1756,Poincare1893,Koyre1952}). The two other primary integrals are given by:
	\begin{align}
    	&\mathcal{P} = m \bf{v}                 \label{eq:01b} \\
    	&\mathcal{M}=\bf{r}\times\mathcal{P}     \label{eq:01c}
	\end{align}    
\end{subequations}

	In order to write down the equations of motion of a given system in a concise way, \Lagrange (\cite{Lagrange1788}) defines a function of the system’s dynamic variables, now known as the Lagrangian $\mathcal{L}$. Its derivatives with respect to time, impulsion, velocity, \textit{etc} $\ldots$ yield its equations of motion. In a closed system or in a central field, total energy $\mathcal{E}$ and Lagrangian $\mathcal{L}$ are univocally linked and (\ref{eq:01a}) can be used to write the full equations of motion of the planet. Equation (\ref{eq:01c}) introduces the kinetic revolution moment $\mathcal{M}$  (in $kg.m^2.s^{-1}$) of planet $m$. In (\ref{eq:01a}) the quantity $\dfrac{\mathcal{M}^{2}}{2mr^{2}}$ is called the centrifugal energy, as opposed to the centripetal energy $\mathcal{U}(r)$. In the case of a revolution with constant $r$ or small eccentricity $e$, the impulsion (or impetus) $\mathcal{P}$ reduces to its angular component: 
\begin{subequations}
    \begin{equation}
        \dot{p}_{\varphi} = mr^{2} \dot{\varphi}         \label{eq:02a}
    \end{equation}

    Given the law that links $\mathcal{M}$  and $\mathcal{P}$  and the conservation of kinetic moment,
\begin{equation}
    \mathcal{M} = mr^{2}\dot{\varphi} = C^{st}     \label{eq:02b}
\end{equation}    

	From a geometrical stand point, $\dfrac{1}{2}r \cdot r d\varphi$ in (\ref{eq:02b}) represents the area $df$ of a sector of the orbit formed by the two infinitesimal vectors and the element of arc of the orbit/trajectory of $m$. Thus, one can write:
\begin{equation}
    \mathcal{M} = 2 m \dot{f}     \label{eq:02c}
\end{equation}    
\end{subequations}

	If one integrates (\ref{eq:02c}) along the full ellipse of revolution with period $T$, (\ref{eq:02c}) becomes:
\begin{subequations}
    \begin{equation}
        2 m f = T \mathcal{M} 
        \label{eq:03a}
    \end{equation}
    
    The surface $f$ is equal to $ab$, where $a$ and $b$ are the semi-major and semi-minor axes of the ellipse. $p$ and $e$ being respectively the ellipse’s parameter and eccentricity, one gets (\textit{eg}. \cite{Landau1988}): 
\begin{equation}
    a = \dfrac{p}{1-e^2} = \dfrac{\alpha}{2 \left\lvert \mathcal{E}\right\rvert}, \quad b = \dfrac{p}{\sqrt{1-e^2}} = \dfrac{\mathcal{M}}{2 m \left\lvert \mathcal{E}\right\rvert}
    \label{eq:03b}
\end{equation}    

We point out at this stage that when $m$ is far from the Sun the semi-major axis depends only on the field, but when it is closer, the displacement takes over. Injecting (\ref{eq:03b}) in (\ref{eq:03a}), one gets:
\begin{equation}
    T = 2\pi a^{3/2}\sqrt{\dfrac{m}{\alpha}} = \pi \alpha \sqrt{\dfrac{m}{2\left\lvert \mathcal{E}\right\rvert}}
    \label{eq:03c}
\end{equation}    

    This is \Kepler’s third law (see \cite{Warrain1942}), $T^2$ proportional to  $a^3$ . The constant ratio  $\dfrac{a^3}{T^2} = K$ is given by Newton’s law:
\begin{equation}
    K = \dfrac{\mathcal{G}M_s}{4\pi^2}
    \label{eq:03d}
\end{equation}    

	with the gravitational constant $\mathcal{G}$=  6.67384 x10$^{-11}$ $m^3.kg^{-1}.s^{-2}$ and the Sun’s mass $M_s$  = 1.98892 x10$^{30}$ kg.    \\

    Using equations (\ref{eq:03}) and the planetary values listed in Table A, one can calculate a value $\mathcal{G}_p$ of $\mathcal{G}$. We have done it in the following way. For a given planet, the Sun’s attraction is given by $\dfrac{\mathcal{G}M_s m}{r^2}$ and derives from the potential $\dfrac{\mathcal{G}M_s m}{r}$ which has the form $\dfrac{\alpha}{r}$ being given by (\ref{eq:03c}). So, for each planet one can write:
\begin{equation}
    \alpha = 4\pi^{2} \dfrac{a^3}{T^2} *m = \mathcal{G}_{p} Ms*m
    \label{eq:03e}
\end{equation}    

and finally:
\begin{equation}
    \mathcal{G}_{p} = \dfrac{4\pi^{2} a^3}{M_s T^2}
    \label{eq:03f}
\end{equation}    
    \label{eq:03}    
\end{subequations}
   
\begin{figure}[!h]
		\centering{\includegraphics[width=\columnwidth]{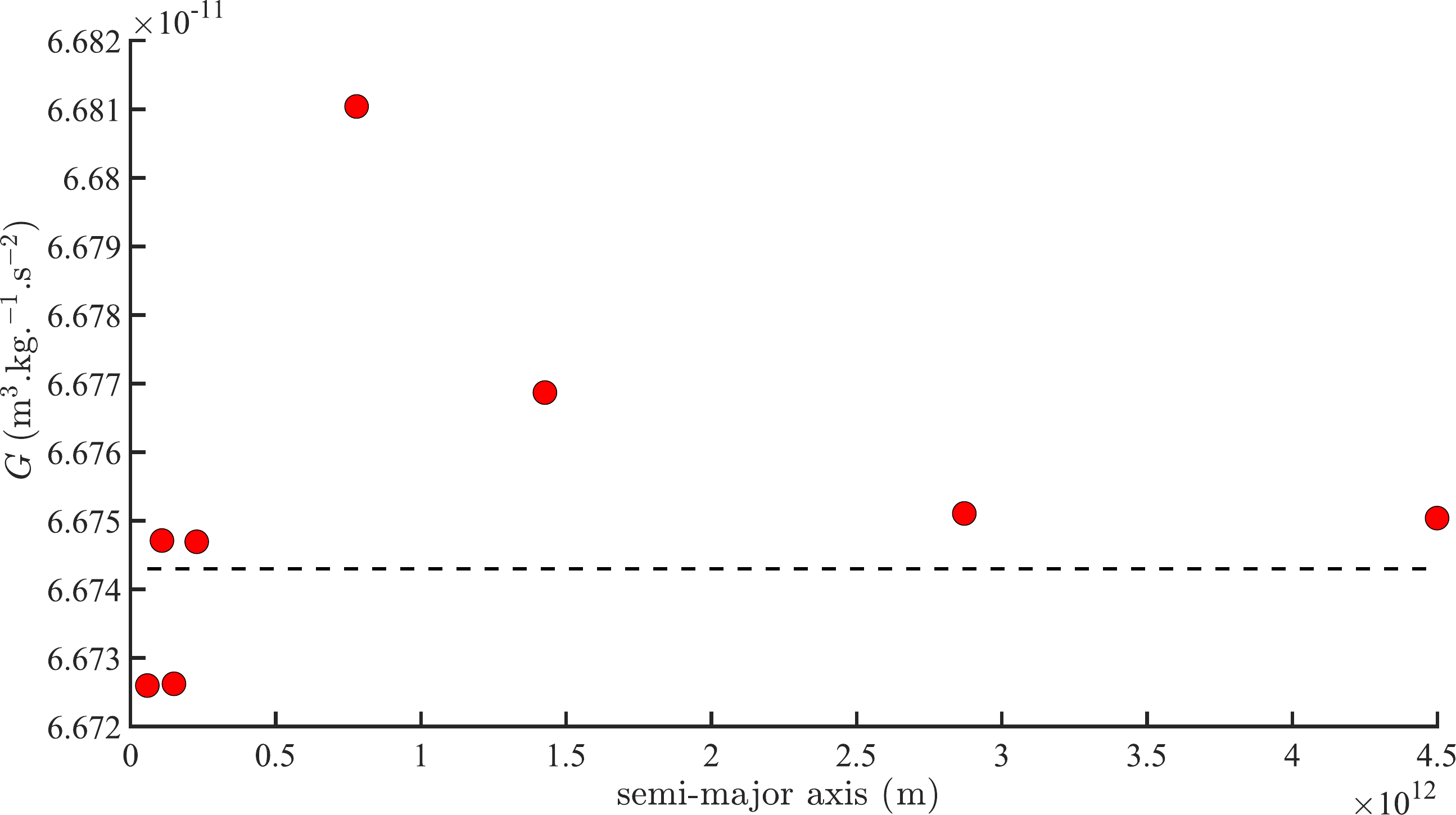}} 	
     	\caption{Gravitational constant $\mathcal{G}_{p}$ calculated for each planet following equation (\ref{eq:03f}), that is using Kepler’s third law for each planet.}
		\label{Fig:01} 	
\end{figure}	   

	Results are shown in Figure \ref{Fig:01}. The first four points correspond to the terrestrial (inner) planets (Mercury, Venus, Earth and Mars). Their $\mathcal{G}_{p}$ values are similar with no specific behavior as a function of distance to the Sun. In contrast, the points corresponding to the giant (outer) planets (Jupiter, Saturn, Uranus and Neptune) follow a regular pattern with $\mathcal{G}_{p}$ decreasing regularly as a function of distance to the Sun (aphelia), and tending towards $\mathcal{G}$ (dashed line). This is readily understood: Kepler’s third law ensures that the ratio $\dfrac{T^2}{a^3}$ be constant, implying that with each full revolution the planet spans the same surface of the ellipse of revolution. But this does not imply that after a full revolution in a central field the planet returns to its initial position in the universe. The trajectory of a revolution is not closed and precession results. In a Galilean system, total energy $\mathcal{E}$ is conserved and the energy corresponding to the central field $\mathcal{U}(r)$ is a constant, and the same for all planets. The balance between the kinetic energy and the centrifugal energy (that takes the moment into account) ensures that the planet retains the same trajectory.
   
	The equation of motion of $m$ can be derived from (\ref{eq:01}). For the radial coordinate one obtains by integration from (\ref{eq:01a}): 
\begin{subequations}
    \begin{equation}
        \dot{r} \equiv \dfrac{dr}{dt} = \sqrt{\dfrac{2}{m} [\mathcal{E}-\mathcal{U}(r)]-\dfrac{\mathcal{M}^{2}}{m^{2}r^{2}}}
        \label{eq:04a}
    \end{equation}
    
    Separating variables and integrating:
    \begin{equation}
           t = \int \dfrac{dr}{\sqrt{\dfrac{2}{m} [\mathcal{E} - \mathcal{U}(r)]-\dfrac{\mathcal{M}^2}{m^{2}r^2}} } + C^{st}
            \label{eq:04b}
    \end{equation}

    From (\ref{eq:02b}) one derives:
    \begin{equation}
            d\varphi = \dfrac{\mathcal{M}}{mr^{2}}dt
            \label{eq:04c}
    \end{equation}
    
    Integrating (\ref{eq:04c}) in (\ref{eq:04a}):
     \begin{equation}
            \varphi = \int \dfrac{\dfrac{\mathcal{M}}{r^2} dr}{\sqrt{2m [\mathcal{E} - \mathcal{U}(r)] - \dfrac{\mathcal{M}^2}{r^2} }}+C^{st}
            \label{eq:04d}
    \end{equation}

    Equations (\ref{eq:04a}), (\ref{eq:04b}) and (\ref{eq:04d}) are the full equations of motion of a planet $m$ in a central field. The second provides the link between $r$ and $\varphi$ . Equation (\ref{eq:01}) shows that the radial part of the motion can be considered as a linear motion in a central field with "effective" potential energy:
      \begin{equation}
            U_{eff} = \mathcal{U}(r) + \dfrac{\mathcal{M}^2}{2mr^2}
            \label{eq:04e}
    \end{equation}
    \label{eq:04}    
\end{subequations}   
   
This equation underlines the reason why the revolution momentum is so important for geophysicists. For instance, one may ask what are the boundaries of  the domain covered by the planet, that is when (\ref{eq:01}) reduces to:
\begin{subequations}
    \begin{equation}
        \mathcal{E} = \dfrac{\mathcal{M}^{2}}{2mr^{2}} +\mathcal{U}(r)  
        \label{eq:05a}
    \end{equation}
    
    Radial velocity (\ref{eq:04a}) vanishes, but tangential velocity (\ref{eq:04c}) does not. At the singular points where (\ref{eq:05a}) holds, the function changes from growing to decreasing (and vice-versa). The domain is bounded by two circles $r_{min}$ and $r_{max}$. The trajectory is finite but not necessarily closed. Let  $\Delta \varphi$ be the angle covered by the planet as $r$ decreases from $r_{max}$ and  $r_{min}$ then grows back to  $r_{max}$. From (\ref{eq:04d}):
    \begin{equation}
        \Delta \varphi = 2 \int_{r_{min}}^{r_{max}} \dfrac{\dfrac{\mathcal{M}}{r^2} dr}{\sqrt{2m [\mathcal{E} - \mathcal{U}(r)] - \dfrac{\mathcal{M}^2}{r^2} }}
        \label{eq:05b}
    \end{equation}
     \label{eq:05}    
\end{subequations}   

	Let $\delta U = \dfrac{\gamma}{r^3}$ be the (small) perturbing potential from a second planet, and let $\mathcal{U}(r) = -\dfrac{\alpha}{r} + \delta U$ , then integrate (\ref{eq:04d}). One gets (recall that $(\textrm{acos}\  u)^{'} = \dfrac{-u^{'}}{\sqrt{1-u^2}}$):
\begin{subequations}
    \begin{equation}
        \varphi = \textrm{acos}  \dfrac{ \dfrac{ \mathcal{M} }{ r } - \dfrac{m\alpha}{\mathcal{M} } } {\sqrt{2m \mathcal{E} - \dfrac{m^{2}\alpha^{2}}{\mathcal{M}^2} }}+C^{st}    
        \label{eq:06a}
    \end{equation}
     
     From (\ref{eq:03b}) one has $p = \dfrac{\mathcal{M}^2}{m\alpha}$ and $e = \sqrt{1+\dfrac{2\mathcal{E}\mathcal{M}^2}{m\alpha^{2}}}$ and (\ref{eq:06a}) can be written:                    
     \begin{equation}
        \dfrac{p}{r} = 1 +e \cos \varphi        
        \label{eq:06b}
    \end{equation}
    
    One can write the perturbing field $r^2 \delta U = \dfrac{\gamma}{r}$. (\ref{eq:05b}) can now be written as:
    \begin{equation}
        \Delta \varphi = -2 \dfrac{\partial}{\partial \mathcal{M}} \int_{r_{min}}^{r_{max}} \sqrt{2m(\mathcal{E}-\mathcal{U}) - \dfrac{\mathcal{M}^2}{r^2}}
        \label{eq:06c}
    \end{equation}
  
    With $\mathcal{U}(r) = -\dfrac{\alpha}{r} + \delta U$ and developing the expression under the integral in successive powers of $\delta U$ , the order 0 term of the displacement is $2\pi$ and the order 1 term is:
    \begin{equation}
        \delta \varphi =  - \dfrac{6\pi \gamma}{\alpha p^2}
        \label{eq:06d}
    \end{equation}

    Given (\ref{eq:03b}) and (\ref{eq:03c}), (\ref{eq:06d}) becomes:
    \begin{equation}
        \delta \varphi = \dfrac{24\pi^{3} a^2}{T^{2} (1-e^{2})} * \dfrac{\gamma}{a \alpha}
        \label{eq:06e}
    \end{equation}
    
    If one writes $\dfrac{\gamma}{a \alpha} = \dfrac{1}{c^2}$ then (\ref{eq:06e}) becomes \Einstein’s formula for the precession of perihelion:
    \begin{equation}
        \delta \varphi_{einstein} = \dfrac{24\pi^{3} a^2}{T^{2} c^{2}(1-e^{2})}  = \dfrac{6\pi M_s \mathcal{G}}{c^{2}a (1-e^2)}
        \label{eq:06f}
    \end{equation}
     \label{eq:06}    
\end{subequations}      
\newpage 
\section{\label{sec3} Distributions of some planetary parameters: power-laws of aphelia}
\begin{figure}[H]
		\centerline{\includegraphics[width=\columnwidth]{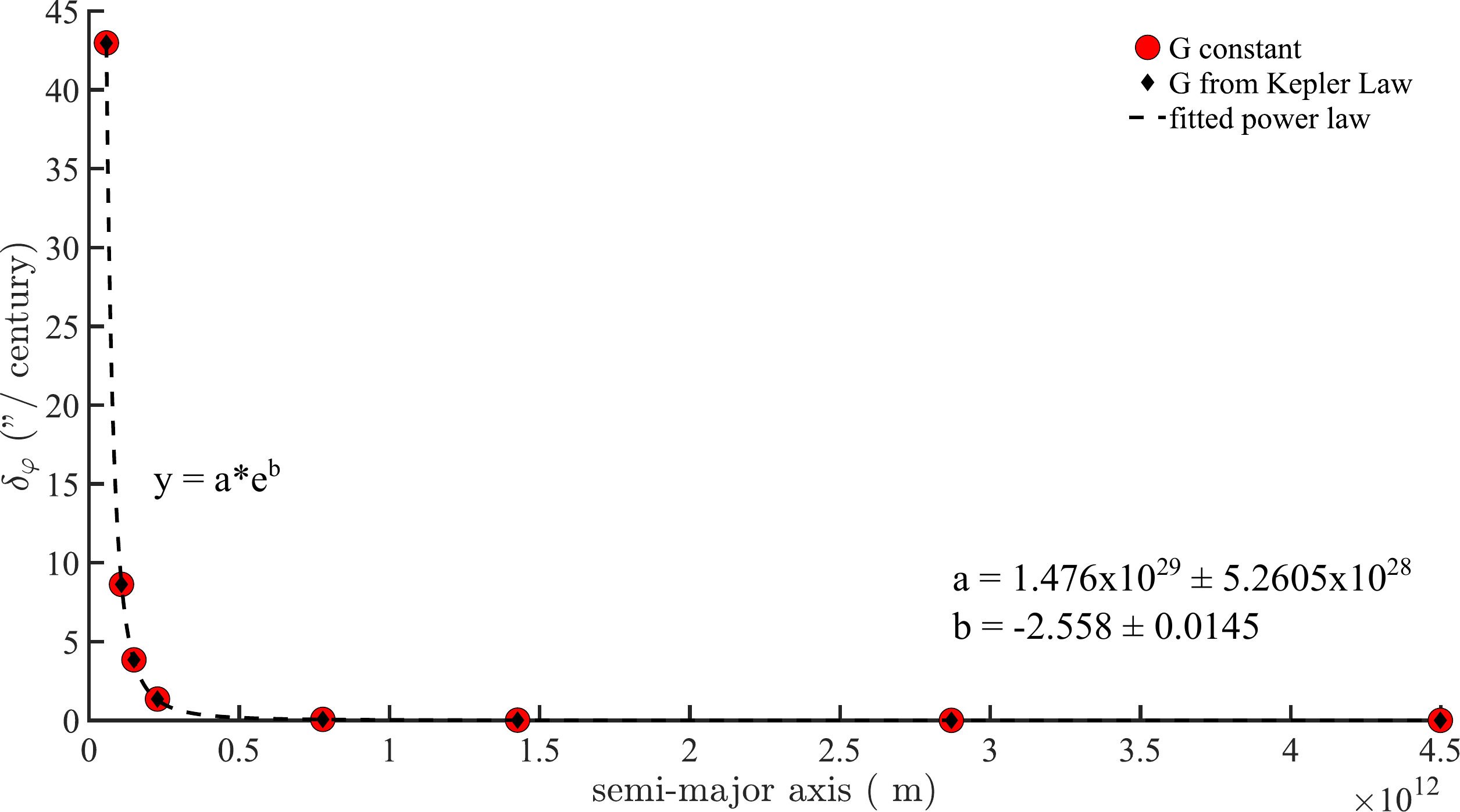}} 	
		\caption{Delay $\delta \varphi$ of the precession of perihelion for solar System planets, using Einstein’s formula (\ref{eq:06f}) and the gravitational constants from Figure \ref{Fig:01}.}
		\label{Fig:02}
\end{figure}

	Parameter $\alpha = mM_{s}\mathcal{G}$  can be fully determined from the values listed in Table A (in Appendix A), through Kepler’s law (\ref{eq:03c}). $\mathcal{G}$ follows Figure \ref{Fig:01}. We next calculate the delay in the planets’ perihelia, using Table \blue{A} (in Appendix \blue{A}) and Einstein’s formula under two cases:
\begin{enumerate}
    \item the values of gravitation (black diamonds) are as in Figure \ref{Fig:02};
    \item one assumes a constant $\mathcal{G}$  (red dots).
\end{enumerate}

\begin{figure}[H]
		\centerline{\includegraphics[width=\columnwidth]{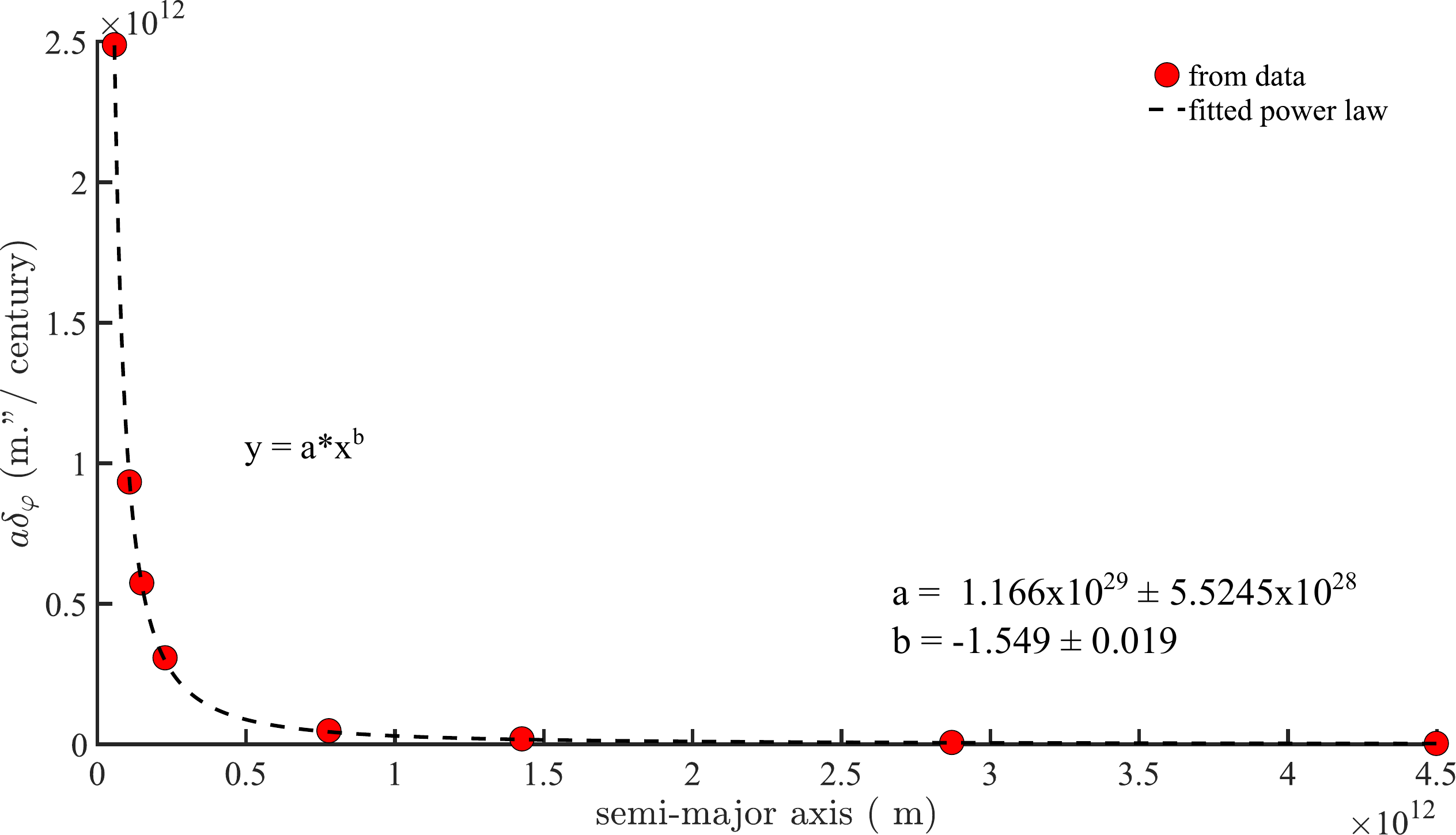}} 	
		\caption{Evolution of the Kepler-Lagrange surface of delay $a\delta \varphi$ (see text).}
	\label{Fig:03}
\end{figure}

	The two determinations are essentially identical as seen in Figure \ref{Fig:03} and a good fit with a power law is shown as a dashed curve. It is noteworthy that the exponent is indistinguishable from (-5/2). This result is readily understood in Lagrange’s formalism. Torques have their origin in the revolution of planets, that act on the Sun. The Sun’s rotation is modified, as evidenced for instance by sunspots. Since the system is Galilean with uniform time, the modified solar rotation acts instantaneously on the rotation axes of planets. Both the classical and the relativistic interpretations are similar, with the two parameters $ 1 /c^{2} $ and $\gamma / (\alpha a) $ as the link. 

	The $r \delta \varphi$ (or $a \delta \varphi$) can be regarded as the apparent surface of delay (or advance) with respect to closed trajectories (\Kepler’s second law). That delay does not depend on time, since $t$ does not appear in equation \ref{eq:06e}.  So $a \delta \varphi$ follows the law of areas and should have dimension $a^{-3/2}$. This is indeed seen in Figure \ref{Fig:03}.
	
Thus $a \delta \varphi$ behaving as $a^{-3/2}$ implies that  $\delta \varphi$ behaves as  $a^{-5/2}$ as found in Figure \ref{Fig:03}. The delay in the periaster is a constant in our system; its value depends only on the distance to the origin of the central field. This was understood by Einstein but may be seen more clearly in the Lagrange formalism behind equation \ref{eq:06e}.
	
\begin{figure}[H]
		\centerline{\includegraphics[width=\columnwidth]{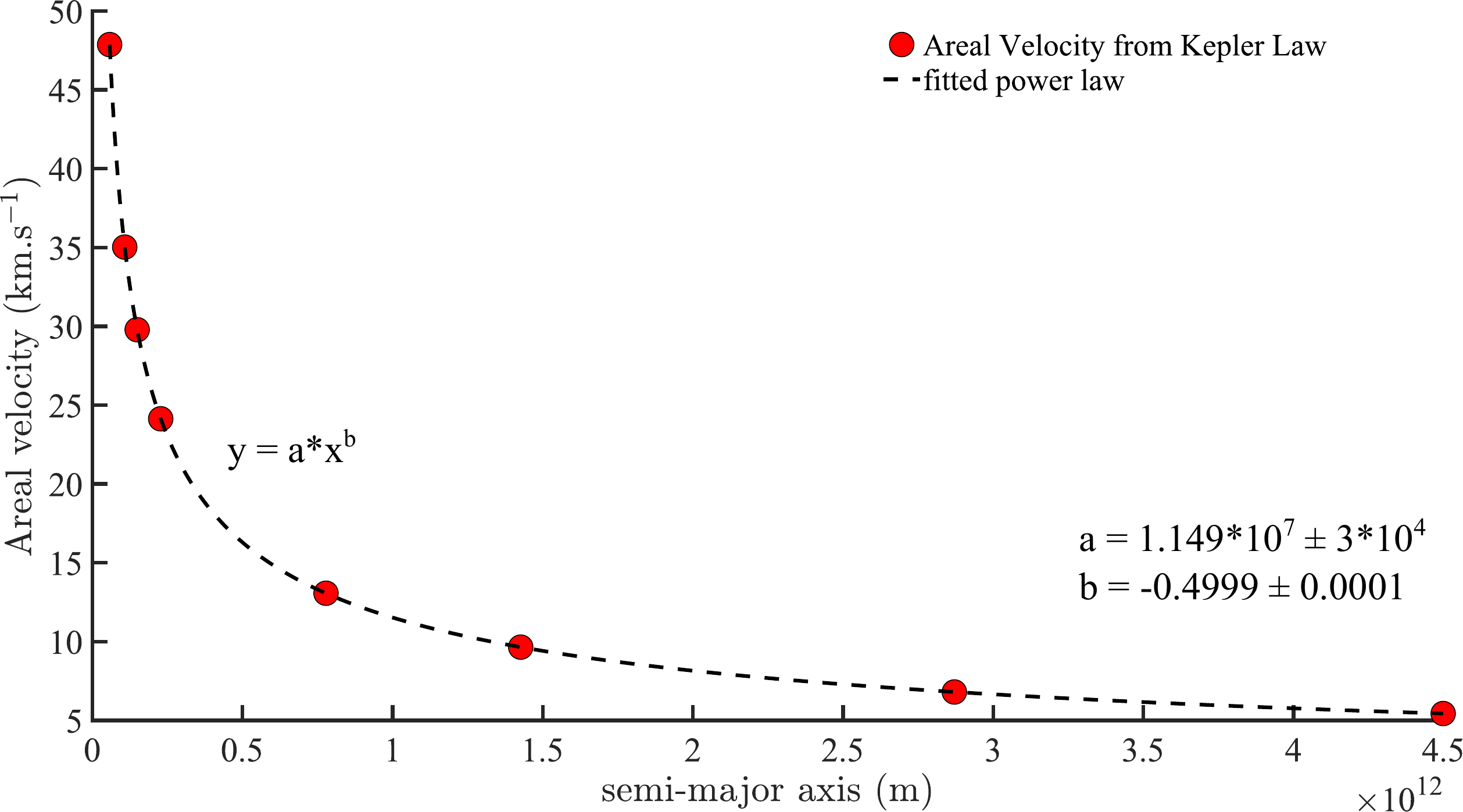}} 	
		\caption{Evolution of areal velocity as a function of planets’ aphelia.}
		\label{Fig:04}
\end{figure}

	Replacing radial coordinate $r$ by aphelia $a$ in equation (\ref{eq:01c}), the centrifugal term, that involves the areal velocity (Figure \ref{Fig:04}) behaves as $a^{-1/2}$.  This is readily understood as, from Kepler’s third law $\dfrac{T^2}{a^3} = C^{st}$, $\dfrac{1}{av^2}$ is dimension-less and $v$ behaves as $a^{-1/2}$.

\begin{figure}[H]
	    \centerline{\includegraphics[width=\columnwidth]{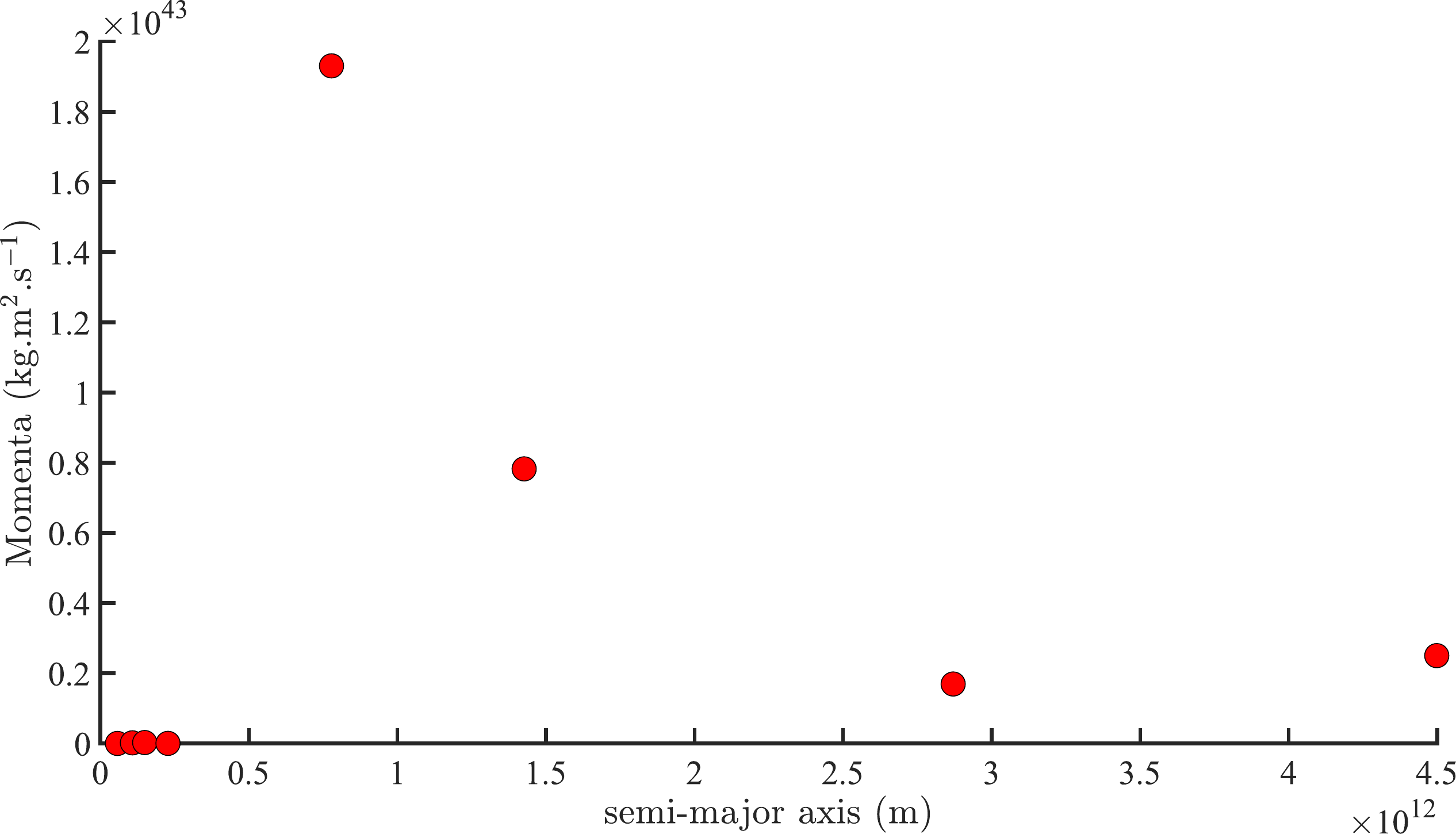}} 	
		\caption{The kinetic moments of the solar system planets as a function of aphelia.}
		\label{Fig:05}
\end{figure}

  Figure \ref{Fig:05} shows the kinetic moments of planets (Appendix A) as a function of distance to the Sun. The moments of telluric planets are essentially negligible whereas those of the giant planets follow a monotonous decreasing trend. It therefore seems that the telluric planets have lost energy. We hypothesize that this energy is transferred to the planets rotation axis. Following \Laplace (\cite{Laplace1799}) and \Lagrange (\cite{Lagrange1788}), any modification of the inclination of the planet’s rotation axis  leads to a modification of the rotation.

	\Laplace (\cite{Laplace1799}) provides the system of linear differential equations that link the derivative of rotation velocity to changes in rotation axis inclination (see \cite{Lopes2021,Lopes2022a}). Geophysical or astronomical perturbations of a planet’s revolution lead to changes in its rotation, We now attempt to identify the corresponding law that is obeyed by the ratio "revolution period/rotation period".

\newpage 
\begin{figure}[H]
		\centering{\includegraphics[width=\columnwidth]{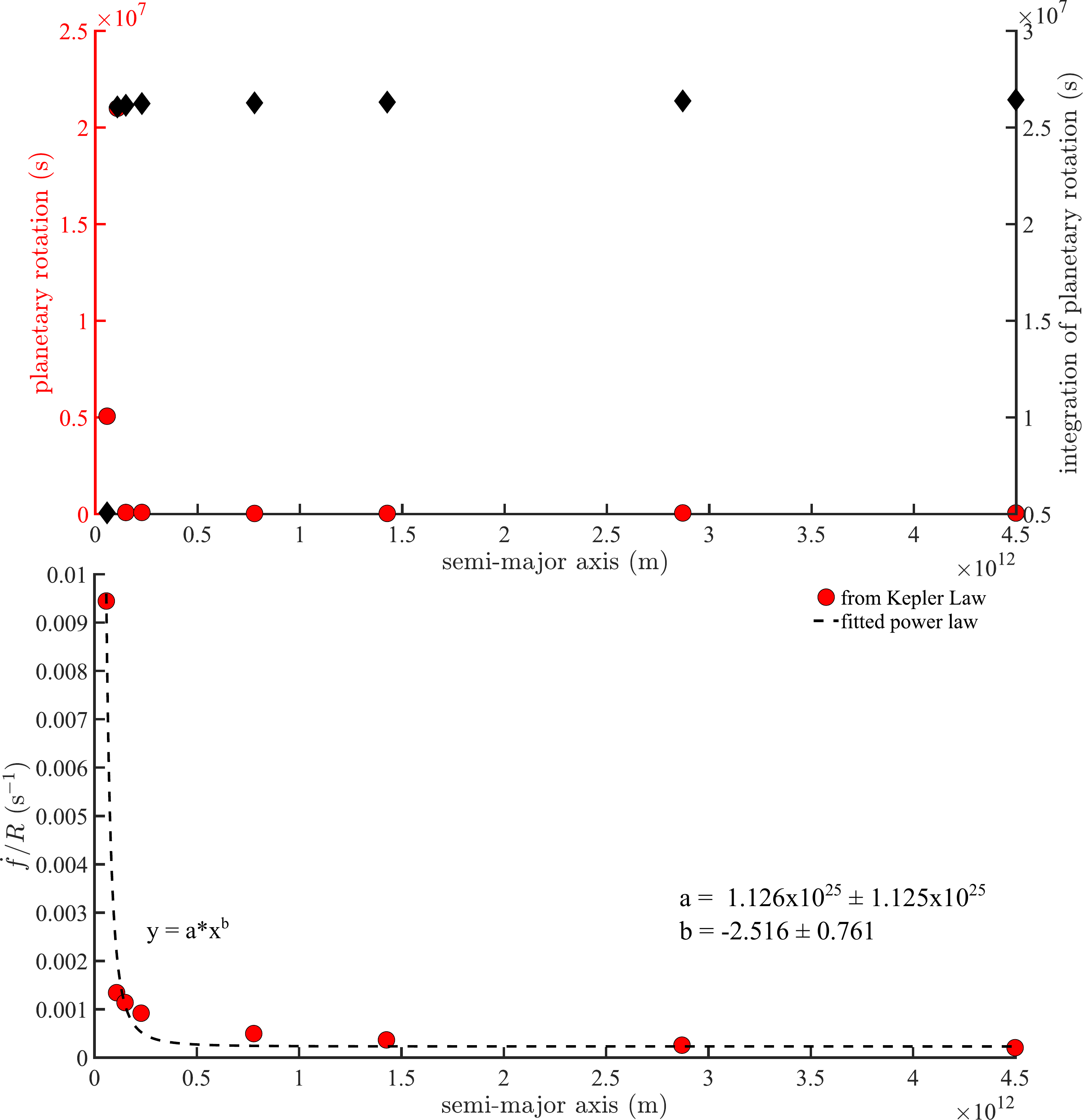}} 	
     	\caption{Red dots = rotation periods of planets as a function of aphelia. Black diamonds = integral of the red dots. Bottom: Evolution of the ratio of areal velocity to the integrated periods (R) on top of figure.}
		\label{Fig:06} 	
\end{figure}	    

\begin{figure}[H]
		\centering{\includegraphics[width=\columnwidth]{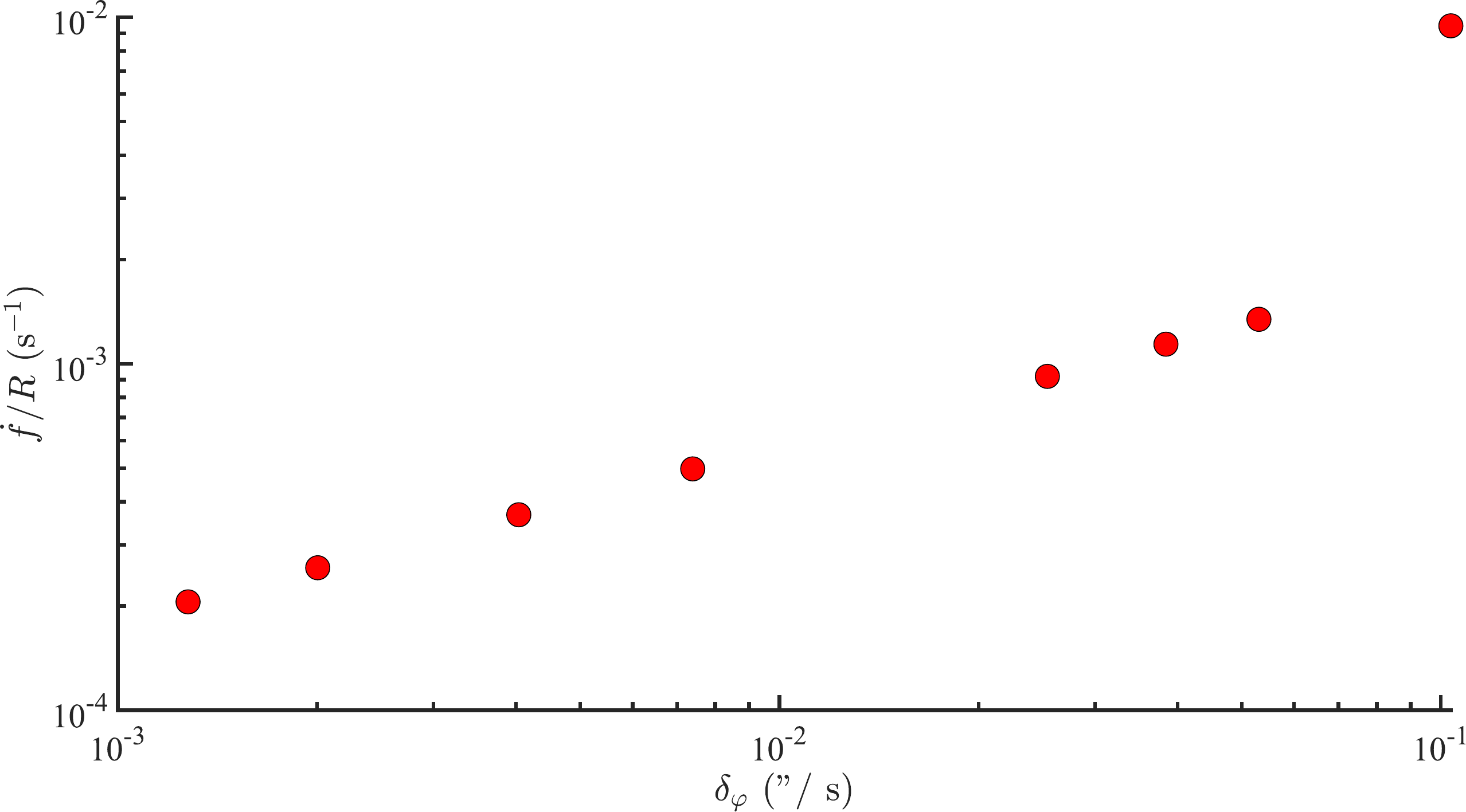}} 	
     	\caption{Evolution of ratio of areal velocity $\dot{f}$ to integrated period R as a function of perihelion delay $\delta \varphi$.}
		\label{Fig:07} 	
\end{figure}

	On the top part of Figure \ref{Fig:06}, red dots represent the rotation periods of the planets (Kepler’s law). One must integrate these rotation periods (Figure \ref{Fig:06}, black diamonds) to show the importance of the moment. The perturbing phenomenon prevents the elliptical trajectory from closing; this is not a “theoretical” statement but primarily an observational constraint, namely the precession of equinoxes. The ratio of areal velocities to rotation gives a $a^{-5/2}$ law (Figure \ref{Fig:02}).  We have chosen the $\dfrac{\gamma}{r^3}$ form for potential $\delta\mathcal{U}$ to show the similarity with general relativity theory (equation \ref{eq:06f}). Figure \ref{Fig:06} bottom shows the ratio $\dfrac{\dot{f}}{R}$ as a function of the perihelion delay $\delta \varphi$. The fit to an $a^{-5/2}$ dependence is again excellent. But then, this implies a strict linearity of the energy exchange between revolution and rotation (and delay at the perihelion). This is indeed the case as seen in Figure \ref{Fig:07} (slope 2.008 in this log-log diagram).  We should in any case expect to find astronomical signals in most terrestrial geophysical phenomena, such as the occurrence of Jupiter’s period of revolution, the commensurable periods of Jupiter and Saturn, or of the precession of equinoxes. And indeed we do.

\section{Discussion and Conclusion} 
	Section \ref{sec2} of this paper has proposed a summary of key results obtained by Lagrange \cite{Lagrange1788} on classical mechanics, a theory better known as the “theory of the spinning top”, that he applied to astronomy. In Section \ref{sec3} we have applied it to show that a number of planetary quantities follow power-law distributions. Equation \ref{eq:01}, the starting point, gives the Lagrangian energy of a planet in motion in a central field. Borrowing an optical analogy from \blue{Hooke} (\cite{Koyre1952}), \Newton focused on the centripetal forces that act on planets gravitating about the Sun. Indeed, centripetal forces are the only ones that appear in Newton’s table of contents. Although this is not widely recognized, some of Newton’s conclusions were criticized by several of his contemporaries and successors (see  \cite{Laplace1799,dAlembert1749,Chastelet1756,Poincare1893}). Rightly so, as his theory did not match some of the observations. In the present paper, we have followed the \textit{lagrangian} approach that actually underlies the equations of most physics papers and books (\eg \cite{Poincare1893,Landau1988,Milankovic1920}).

	The Lagrange integrals are "prime" integrals, that is they preserve physical quantities when the system is either closed or open but with a central field $\mathcal{U}(r)$, with a single symmetry axis (\cite{Lagrange1788}). The most important equations are (\ref{eq:01a}, \ref{eq:01b}, \ref{eq:01c}). They apply in a Galilean reference, for a physical system that is in that case open with a central field. As we have seen, that central field $\mathcal{U}(r)$ is perturbed. The dependence is not any more an exact $\dfrac{1}{r}$ dependence, therefore the orbits are not closed and precession takes place. Fortunately, since this perturbation is very small, one can use Lagrangian mechanics throughout.

	As a counter-example, the conservation of kinetic momentum does not apply to an elastic Earth (\cf \blue{Lambeck}  \cite{Lambeck2005}, chapter 3). Indeed, most researchers who discuss an elastic Earth consider a single isolated planet without any central field. In other words, the kinetic moment, which is linked to the order 2 inertia tensor, may not be symmetrical and diagonalizable any more. This is already clearly stated by \Laplace (\cite{Laplace1799}) and \Poincare (\cite{Poincare1893}). The equations are valid for a solid Earth or one in which deformations are negligible. In equation (\ref{eq:01a}), there is a competition between 3 energies, kinetic, centrifugal and gravitational attraction (the last one is attractive or centripetal, and identical for all planets). The farthest the distance to the Sun, the larger the influence of  $\mathcal{U}(r)$. The only planet that does not fit the overall linear law as seen in Figure \ref{Fig:07} is Neptune.

	The important law, and an \textit{observational} one, established by \Kepler in 1619 (\cf \cite{Warrain1942}) is that the ratio of the square of the period of revolution to the cube of aphelia is a constant. A planet can revolve about the Sun without coming back to its initial location in the universe. That is the phenomenon of precession  (\eg \cite{dAlembert1749, Milankovic1920,Lopes2022b}).We have seen that the conditions for trajectory closure are not met. Perihelion can be ahead of time or delayed with respect to the prediction by Newton. For instance, in 1869, Urbain Le Verrier could not find the delay of Mercury despite including attractive forces from all planets (\eg \cite{Weinberg1972}). In an attempt to circumvent this problem, several authors included the flattening of the Sun in the attractive forces (\eg \cite{Newcomb1882}). The discrepancy between Newtonian theory and observations has led to numerous studies on the possible variations of the gravitational "constant", particularly since the 1970s (\cite{Pochoda1964,Shapiro1971,Wu1986,Combes2010,Alvey2020}).

	In equation (\ref{eq:01a}), we have seen that $(\dfrac{dr}{dt})^2$ can be zero without the planets being motionless (the trajectories become circles). In one of the two remaining terms in equation (\ref{eq:01a}), there is one, centrifugal energy, that is planet-dependent. All gravitating planets precess according to a (-5/2) power law of aphelia (Figure \ref{Fig:02}). We have sought a small perturbation of $\mathcal{U}(r)$  that would prevent trajectories from closing, with a dependence on aphelia that is of a lesser degree than the centrifugal force (or planets would leave their finite trajectory). The function $\dfrac{1}{r^3}$ meets these constraints ($\gamma$,$\alpha$). The Lagrange and Einstein theories then lead to the same result. For Lagrange the perturbation $\delta \mathcal{U}$ could only be due to the interactions of torques, since $\mathcal{U}$ is a constant imposed by the mass and immobility of our star, and $(\dfrac{dr}{dt})^2$ can be zero without changing drastically the astronomical orbits. The only remaining term is the centrifugal force due to planetary rotations, which amounts to torques.

	In Figure \ref{Fig:05}, we see that there is a difference in amplitude and behavior of the $\mathcal{G}$ and kinetic moments of planets, both quantities being observations. Yet, the perihelion delays (Figure \ref{Fig:02}), the areal velocities (Figure \ref{Fig:04}) and the planetary rotations (Figure \ref{Fig:06}, top) display power laws of aphelia, whose behavior contrasts for instance with that of the kinetic moment (Figure \ref{Fig:05}). The areal velocity being linearly linked to the kinetic moment of planets (equation \ref{eq:02c}), this must be the level at which the transfer is achieved.  Given the instantaneity and reciprocity of Galilean systems, any torque acting on the Sun is returned to Earth, whose rotation axis is perturbed. If one considers the law of rotation periods as a function of aphelia, we obtain – to a small error – the variations of inclination of the rotation axis. 
	
	An important and useful relation is illustrated by Figure \ref{Fig:06} (bottom) that shows the power law linking for all planets their areal velocity to their integrated period, \ie the ratio $\dfrac{\dot{f}}{R}$. Since we are in a closed system, this ratio must be constant for each planet, and all planets do follow the same power law of aphelia ($\mathcal{M}$ being actually ”hidden” in the transfer). We have seen (equation \ref{eq:01a}) that the Lagrangian energy of a planet consists in three energies whose sum must be conserved. There is no source of friction in the universe and the trajectories of planets have remained stable for long times (we do not refer here to the much longer durations where chaotic behavior sets in). So, how is the energy surplus ($\delta \mathcal{U}$) dissipated without altering the planetary orbits? This dissipation could occur through viscous flow, elasticity, friction $\ldots$ But then the effects of this ($\delta \mathcal{U}$) should be visible in actual observations. And such is indeed the case for geophysical and climatic phenomena recorded in terrestrial observatories. The oldest fields of study in geophysics, that is geomagnetism, polar motion and fluid motions (sea-level, atmosphere), provide the longest series of observations. The signatures of planetary effects have been identified for a long time, not only in geophysical observations \cite{Morth1979,Morner1984,Fairbridge1984,Lopes2017,Dumont2020,Scafetta2020,Dumont2021, Lemouel2021, Bank2022, Courtillot2022, Dumont2022,Lopes2023} but also in heliophysical series such as sunspots (\eg \cite{Bartels1932,Abreu2012,Lemouel2020,Courtillot2021}). The origins of many of these signatures are still hotly debated.
	
	The small energy term $\delta \mathcal{U}$ in \Lagrange’s formulation (\cite{Lagrange1788}) influences the Earth’s rotation axis (as in the \Lagrange top), leading to the consequences predicted by Laplace \cite{Laplace1799}. Then, given \Laplace’s equations (\eg \cite{Lopes2021,Lopes2022a}) that are close to the famous \blue{Liouville-Euler} equations (\eg \cite{Lambeck2005}, chapter 3), both the external and internal mobile masses will be forced to re-organize not only due to the luni-solar torques but also to the planetary torques as proposed by \Laplace (\cite{Laplace1799}).
	
	In closing, as illustrations of the formulation of Lagrange’s mechanics, we address two significant geophysical questions.
	
	First, why has Earth rotation accelerated since 2020 (\eg \cite{Trofimov2021}) ? It has been shown (\eg  \cite{Laplace1799,Poincare1893,Lopes2021,Lopes2022a}) that analysis of the length of day (\textit{lod}), directly linked to the rotation velocity of Earth, and analysis of polar motion (rotation) lead to the same results (\eg \cite{Laplace1799, Lopes2022a}). If the Earth does behave as a spinning top, the lod is affected by changes of the inclination of the rotation axis. Laplace shows that the period of the \blue{Chandler} free oscillation (\cf \cite{Chandler1891a,Chandler1891b})(the \blue{Euler} oscillation, a function of the axial and equatorial moments of our planet) ranges between 306 and 578 days. Laplace concludes that this is due to variations in the inclination of the rotation axis. We have analyzed polar motion (\eg \cite{Lopes2017,Lopes2021}) and indeed it is composed of the sum of all periods and combinations of periods of both the telluric and Jovian planets.

	A second question is why the periodicities of the geomagnetic field are so close to those of sunspots (\eg \cite{Bartels1932}). Analysis of the time series of the \textit{Dst} and \textit{aa} geomagnetic indices (\eg \cite{LeMouel2019b}) shows that they follow a \blue{Kolmogorov} power law (\cite{Kolmogorov1941}) with exponent (-5/3).  This comes from direct measurements made since the beginning of the XX$^{th}$ century. It confirms an earlier compilation (\cf \blue{Courtillot and Le Mouël} \cite{Courtillot1984}, figure 45), in which geomagnetic, archeomagnetic and paleomagnetic variations encompassing the past 1 million years also obeyed a Kolmogorov power law with exponent (-5/3). In a series of papers, \LeMouel \cite{LeMouel1984}, \blue{Jault and Le Mouël} (\cite{Jault1991}) and \blue{Le Mouël et \textit{al.}} (\cite{LeMouel2023}) noted that the secular variation of the geomagnetic field correlates rather well with lod and polar motion. These authors hypothesized a dynamo mechanism where core fluid formed a cylinder tangential to the inner core, forced through an exchange of angular moment by variations in the Earth’s rotation axis. But the order of magnitude of the coupling to the core-mantle boundary was found to be too weak to allow transfer of moment from the mantle to the core.

	If one accepts that the large planetary torques (mainly from the jovian planets) listed in Table A can drive the Earth’s dynamo, as well as sunspots, then these turbulent flows of incompressible fluids that are constantly fed energy by the planetary torques (in the Sun as well as in the Earth’s core) meet the two conditions stated by Kolmogorov \cite{Kolmogorov1941} for his famous (-5/3) power law. Such a law does apply to sunspots (\eg \cite{Lemouel2020}). Using filter theory and (only) the ephemerids of the jovian planets, we have published a prediction of sunspot cycle 25 (\cf \cite{Courtillot2021}) that remains good to 1\% at the time of writing this paper (early 2023).

	In this paper, we have attempted to summarize \Lagrange ’s formulation of mechanics (\cite{Lagrange1788}) and coupled it to \Laplace ’s (\cite{Laplace1799}). Thus, we have shown that it was mechanically (physically) possible to propose that planetary moments could be the main driver of a number of geophysical and heliophysical phenomena, several of which have been monitored for up to three centuries. 

\appendix

\section{A summary of some planetary constants}
\begin{figure}[H]
		\centering{\includegraphics[width=1.1\columnwidth]{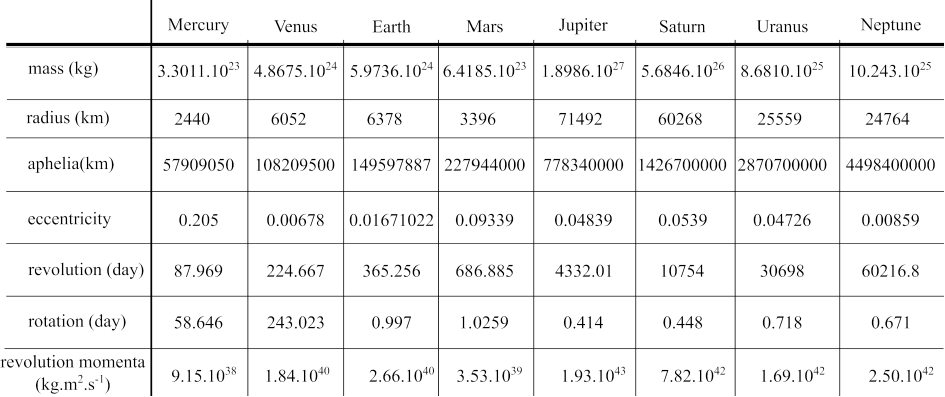}} 
		\label{AppendixA}	
\end{figure}

\bibliographystyle{ieeetr} 
\bibliography{lagrange_biblio}
\end{document}